\documentclass[aps,prd,twocolumn,superscriptaddress,nofootinbib]{revtex4-2}

\usepackage{amsmath,amssymb,graphicx,hyperref,xcolor}
\usepackage{multirow}
\usepackage{physics}
\usepackage{siunitx}
\usepackage{comment}
\usepackage{acronym}
\usepackage[normalem]{ulem}
\usepackage[f]{esvect}
\usepackage{makecell}
\usepackage{orcidlink}
\allowdisplaybreaks[1]

\newcommand{\JGSU}{Department of Physics,  Key Laboratory of Energy Conversion Optoelectronic Functional Materials of Jiangxi Education Institutes, Jinggangshan University, Ji'an 343009, China}
\newcommand{\IAA}{Institute for Astronomy and Astrophysics, Department of Physics, JingGangShan University, Ji’an, 343009, China}
\newcommand{\MAE}{Jiangxi Provincial Key Laboratory of Modern Agricultural Equipment, Ji'an, 343009, China}
\newcommand{\USTC}{Department of Astronomy, University of Science and Technology of China, Hefei, Anhui 230026, China}
\newcommand{\YNU}{School of Physics and Astronomy, Key Laboratory of Astroparticle Physics of Yunnan Province, Yunnan University, Kunming, 650091, China}
\newcommand{\YXU}{Department of Physics, Yuxi Normal University, Yuxi, 653100, China}

\newcommand\XMM{XMM-Newton }

\hypersetup{
    colorlinks = true,
    linkcolor = blue,
    citecolor = blue,
    urlcolor  = blue
}

\begin{document}

\title{X-ray Fourier lag-frequency spectra modulated by stochastic turbulent acceleration in the jets of high-frequency-peaked BL Lac}

\author{Guang-Cheng Xiao}
\affiliation{\JGSU}
\affiliation{\IAA}
\author{Wen Hu}
\email{Corresponding author: huwen.3000@jgsu.edu.cn}
\affiliation{\JGSU}
\affiliation{\IAA}
\affiliation{\MAE}
\author{Da-Guo Jiang}
\affiliation{\JGSU}
\author{Jun-Xian Wang}
\affiliation{\USTC}
\author{Zhen-Yi Cai}
\affiliation{\USTC}
\author{Da-Hai Yan}
\affiliation{\YNU}
\author{Fang-Wu Lu}
\affiliation{\YXU}
             
\date{\today}

\begin{abstract}
X-ray interband time lags are key diagnostics of jet physics and are frequently detected in high-frequency peaked BL Lac (HBL) objects at different epochs with various X-ray telescopes. 
In this work, we theoretically investigate Fourier lag-frequency spectra using a generic one-zone leptonic model incorporating the stochastic turbulent acceleration (STA), 
which plays a crucial role in shaping the emitted photon spectra.   
We demonstrate that the competition between STA, radiative cooling, and escape processes not only gives rise to two well-defined time-lag regimes: hard/positive and soft/negative lags, 
but also reveals the existence of a transition between the two regimes.
Our results indicate that time lags in the transitional and soft-lag regimes can be clearly amplified and modified by STA's suppression of high-energy electron cooling,
and nonlinear synchrotron self-Compton (SSC) cooling can further amplify the emergence of time lags.
We conclude that the adopted model offers a unifying quantitative framework for interpreting the diverse time-lag signatures observed in the X-ray flares of HBLs. Additionally, SSC cooling effects can account for the
relatively large lags observed in TeV-bright flares, as well as the observed trend between lag amplitude and flare duration: the larger the flare duration, the larger the lag.
\end{abstract}

\maketitle

\acrodef{PSD}{power spectral densities}
\acrodefplural{PSD}{power spectral densities}

\section{Introduction}

Stochastic turbulent acceleration (STA), driven by stochastic interactions with turbulent fluctuations or various plasma waves, is believed to
play an important role in the energization of nonthermal particles in blazars with a highly relativistic jet that points closely toward the Earth and emits large amounts of non-thermal continuum emission
covering the entire electromagnetic spectrum from radio waves to $\gamma$-rays \citep{Urry1995PASP}.
Due to relativistic beaming, the jet radiation outshines any other radiative component associated with the active nucleus \citep{bottcher2002ApJ,Ghisellini2008MNRAS}.

High-frequency peaked BL Lac (HBL) objects are the most extreme subclass of blazars.
The jets can offer an energetically extreme plasma environment that is capable of energizing particles up to TeV energies. 
The shape of the characteristic spectral energy distribution (SED) exhibits a double-hump structure:
 the low-energy hump peaking in ultraviolet/X-ray bands and the high-energy hump peaking in the GeV to TeV $\gamma$-ray range \citep{Abdo2010ApJ}. 
Moreover, the variability is most dramatic and occurs on timescales as short as a few hours or less at the high-energy X-ray/$\gamma$-ray bands \citep[e.g.,][]{Aharonian2009A&A,Abramowski2012A&A,MAGIC2024Mn}.
The primary mechanism responsible for the low-energy hump is believed to be synchrotron emission from relativistic electrons accelerated in the magnetized jet.
The high-energy hump is commonly interpreted as due to the inverse Compton (IC) scattering of synchrotron photons 
from the same electron population \citep[synchrotron self-Compton; e.g.][]{Mastichiadis1997A&A}, though its origin is not yet fully understood.

For HBLs, electrons responsible for the X-rays have the highest energy and, therefore, the fastest cooling timescale.
The underlying particle acceleration process can be tracked more closely by X-rays than by photons in other wave bands.
Many studies have demonstrated that the X-ray spectral shape of the sources is better represented
by a logarithmic parabola rather than by either a single power law or a broken power-law model \citep[e.g.,][]{Massaro2004A&Aa,Massaro2004A&A,Pandey2018ApJ,Devanand2025ApJS}, 
indicating that the X-ray spectra continuously steepen with increasing energies.
Meanwhile, the SED peak energy is anti-correlated with the curvature parameter around the peak.
These are consistent with the predictions of the STA mechanism for the emitting electrons \citep{Katarzy2006A&A,Becker2006,Stawarz2008,Tramacere2011ApJ}.

Particle acceleration in the magnetized plasma jets can power the bright X-ray/TeV $\gamma$-ray flares observed in HBLs such as Mrk 421 \citep[e.g.,][]{Yan2013ApJ,Dmytriiev2021MNRAS}.
Measuring the time-resolved spectral variations, correlations, and possible time lags of X-ray variability in different energy bands can give direct clues on the underlying particle acceleration and cooling processes,
and put tight constraints on the physical properties of the jet  \citep[e.g.,][]{Zhang2002MNRAS,Zhang2006ApJb,Zhang2006ApJ,Abeysekara2017ApJ,Devanand2022ApJ,Gokus2024MNRAS}.
Alternatively, the frequency-resolved time lag analysis contains more detailed information about the characteristic timescales for particle acceleration, energy losses, escape, and light crossing in the emission region \citep{Finke2014ApJ,Finke2015ApJ,Lewis2016,Hu2024ApJ,Hu2025A&A}.

To extract the physical information contained in the Fourier transform-related data products including time lags and power spectral densities (PSDs),
 \cite{Finke2014ApJ,Finke2015ApJ} developed a theoretical model based on solving the Fourier-transformed electron transport equation. 
However, the model was only able to produce soft time lags.
In order to explain the hard X-ray time lags observed in blazars, 
\cite{Lewis2016} further extended the approach by focusing on the generalized transport equation that takes into account particle acceleration.

We note that to explore the global properties of the Fourier transform-related quantities, the electron transport equation in the Fourier domain is solved analytically, 
 and therefore SSC cooling is not included, since it is inherently nonlinear.
 However, the effects of the nonlinear SSC cooling play an important role in determining the blazar light curves (LCs) and energy spectra
\citep{Schlickeiser2009mn,Zacharias2010AA,Zacharias2012MN,Zacharias2012ApJ,Zacharias2013ApJ}.

The purpose of this work is to highlight the impact of STA on the Fourier-frequency-dependent time lags while accounting for the nonlinear SSC cooling effects, 
based on the numerical solution to the kinetic equation governing the intrinsic temporal evolution of electrons radiating in a magnetized plasma blob embedded in the jet.
The paper is structured as follows. 
In Section~\ref{sec:model}, we provide a brief description of our model, and the results are presented in Section~\ref{sec:result}. 
Section~\ref{sec:discu} discusses our findings and summarizes this work.

\section{Time delay model} \label{sec:model}

Generally, the entire nonthermal electromagnetic emission of blazars is widely attributed to the superposition of radiation from various regions along the jets.
For HBLs, their X-ray emission is likely primarily associated with jet physics occurring in the innermost jet region.
We therefore adopt the standard SSC scenario and simplify the geometry to focus on X-ray flares arising from a spatially homogeneous, optically thin spherical plasma blob
pervaded by a tangled magnetic field of uniform strength $B'$, and filled with a population of isotropic electrons described by  $N'_{\rm e}(\gamma',t')$.
The blob travels with a Lorentz factor $\Gamma$ along the jet, which is directed at an angle $\theta_{\rm obs}\sim1/\Gamma$ with respect to the line of sight. 
Due to the blob’s relativistic motion, the synchrotron and SSC radiation emitted by the relativistic electrons in the blob is Doppler boosted in the observer’s frame.

Throughout the paper, primes denote quantities measured in the comoving frame, while quantities in the stationary AGN frame are unprimed, unless specified otherwise.

\subsection{Description of the electron distribution}

In this work, we focus on an interesting acceleration scenario with a clear physical picture, although the real acceleration process may be very complex.
In the picture, a population of relativistic electrons with a power-law energy distribution formed by an efficient acceleration process is injected uniformly throughout the blob.
Subsequently, the injected electrons are re-accelerated by turbulent plasma waves uniformly throughout the emitting region,
and cool through radiation before finally escaping from the system \citep{Yan2013ApJ,Diltz2014TimeDL,Kundu2021ApJ,Davis2022MNRAS}. 
To keep track of the electron population within the plasma blob over time, we can solve the Fokker–Planck (FP) equation including electron injection, escape, acceleration, and energy loss terms.
The FP equation is expressed as:
\begin{eqnarray}\label{acc-transport1}
\frac{\partial N'_{\rm e}(\gamma',t')}{\partial t'} &=& \frac{\partial}{\partial \gamma'} \left\{\gamma'^2D_p(\gamma')  \frac{\partial}{\partial \gamma'}\left[\frac{N'_{\rm e}(\gamma',t')}{\gamma'^2}\right]  \right\} \\
&-& \frac{\partial}{\partial \gamma'}\left[S\left(\gamma'\right) N'_{\rm e}(\gamma',t') \right] 
- \frac{N'_{\rm e} (\gamma',t')}{t'_{\rm esc}(\gamma')} + Q'_{\rm e}(\gamma',t'),  \nonumber
\end{eqnarray} 
where $D_{\rm p}(\gamma')$ is the diffusion coefficient in energy space, $S(\gamma')=-C(\gamma')+A(\gamma')$ is an energy advection term accounting for systematic energy loss ($C$) and/or gain ($A$), 
$t'_{\rm esc}(\gamma')$ is the escape timescale describing the electron escape process, and $Q'_{\rm e}(\gamma',t')$ is a source term that accounts for the electron injection process. 
Note that we neglect the systematic energy gain $A$ in the energy advection term $S$.

Due to the interactions of electrons with the turbulent magnetized medium, the diffusion coefficient can be approximated as a power-law function, $D_{\rm p}=D_0\gamma'^q$, 
with $q=2$ for “hard-sphere” turbulence, $q=5/3$ for Kolmogorov turbulence, and $q=3/2$ for Kraichnan turbulence.
In this work, “hard-sphere” turbulence is chosen \footnote{ 
For resonant scattering by the Alfv\'{e}n waves, the index $q= 2$ is supported by the numerical simulations of freely decaying MHD turbulence \citep{2001PhRvE..64e6405C,2015PhRvL.114g5001B}. Moreover, this index is also consistent with the nonresonant acceleration \citep{2019ApJ...877...71T,2014ApJ...791...71L,2020PhRvD.102b3003D}.}
The acceleration timescale therefore becomes a constant, expressed as $t'_{\rm D}=(4D_0)^{-1}$ \citep{Becker2006}.
Meanwhile, the escape timescale due to the spatial transport of electrons is also independent of electron energy 
based on the relation between the momentum diffusion coefficient and the spatial diffusion coefficient.
Here, the energy-independent escape timescale in the blob is parameterized through the relation $t'_{\rm esc}(\gamma')=t'_{\rm esc}=t'_{\rm D}/\eta'_{\rm esc}$,
where $\eta'_{\rm esc}$ determines the escape efficiency, which is related to the spectral index of the steady-state electron energy distribution \citep{Hu2023ApJ}.

The radiative cooling rate of the accelerated electrons due to synchrotron and SSC processes can be written as $C(\gamma')=(b_{\rm syn}+b_{\rm ssc})\gamma'^2$,
with
\begin{eqnarray}
b_{\rm syn}&=&\frac{4c\sigma_{\rm T}}{3m_ec^2}U'_{\rm B},\\
b_{\rm ssc}&=&\frac{4c\sigma_{\rm T}}{3m_ec^2}\int_0^\infty u'_{\rm syn}(\epsilon')f_{\rm kn}(\epsilon',\gamma')d\epsilon',
\end{eqnarray}
where $m_{\rm e}$ is the rest electron mass, $c$ is the speed of light, $\sigma_{\rm T}$ is the Thomson cross-section, $U_{\rm B}'\equiv B'^2/8\pi$ denotes the energy density of the magnetic field, 
$u_{\rm syn}'(\epsilon')$ denotes the energy density of the synchrotron radiation field,
and the function $f_{\rm kn}(\epsilon',\gamma')$ denotes the integral of the Compton kernel, fully taking into account Klein-Nishina (KN) effects for an isotropic seed photon field \citep{Jones1968,RevModPhys.42.237}.
To reduce the computation time significantly, we adopt the analytical approximations proposed by \cite{Moderski2005},
where the authors express the electron energy as $\tilde{b}\equiv4\epsilon'\gamma'$ with $\tilde{b}=1$ denoting the transition between the Thomson and KN scattering regimes. 
For $\tilde{b}\ll1$, $f_{\rm kn}\simeq1$; for $\tilde{b}\gg1$, $f_{\rm kn}\simeq9(\ln\tilde{b}-11/6)/(2\tilde{b}^2)$.
For $\tilde{b}\leq10^4$, the kernel can be approximated as $f_{\rm kn}\simeq(1+\tilde{b})^{-1.5}$.

For the injection term, we assume that electrons are injected with a power-law distribution in energy, given by
$Q'_{\rm e}(\gamma')=Q_0'\gamma'^{-p} H(\gamma';\gamma'_{\rm i, min},\gamma_{\rm i, max}')$, 
where $H(\cdot)$ is the Heaviside function defined by $H=1$ if $\gamma'_{\rm i, min}\le\gamma'\le\gamma_{\rm i, max}'$ and $H = 0$ otherwise.
Here, $Q'_0, \gamma'_{\rm i, min}, \gamma_{\rm i, max}'$ and $p$ are the injection rate, the minimum Lorentz factor, the maximum Lorentz factor and the power-law index, respectively.

To gain some basic knowledge of how the re-acceleration process of relativistic electrons affects the time lags between two LCs at different bands in X-rays, 
we focus on the dynamics of a single-injection event lasting $t'_{\rm inj}$ and the corresponding evolution of electron and photon distributions.
The injection may arise from a highly efficient physical process, either due to the formation of shocks, 
or the annihilation of magnetic field lines at magnetic reconnection sites \citep[e.g.,][]{Cerutti2012ApJ,Comisso2018PhRvL,Petropoulou2019ApJ,Marcowith2020LRCA}.
 However, its origin is still unclear and is beyond the scope of this work.

Using the full time-dependent electron distribution calculated numerically by solving the kinetic equation \eqref{acc-transport1}, 
the time-dependent emission spectra can be calculated using the standard formulas reported in \cite{Dermer2009ApJ}, which provide the synchrotron and IC emission from a given electron distribution.
Then, the theoretical LCs at different bands can be calculated by integrating over the corresponding range $[E_l, E_u]$.
Any variability on a timescale shorter than the light-crossing time of the system with size $R'$ should not be detectable in principle. 
To avoid causality violations, the light-crossing effect is incorporated into the calculations.
For the calculation of this effect, the emitting region is divided orthogonally to the line of sight into a number of slices each with a thickness $d\ell$ corresponding to the light-travel distance for one numerical time step. 
 The observed flux $\mathcal{F}_{\rm obs}$ at a given observer time $t_{\rm obs}$ is then the volume-weighted sum of the contributions from different slices evaluated at their corresponding retarded intrinsic times $t_{\rm obs}\delta_D/(1+z)-(\ell/c) $: 
 $\mathcal{F}_{\rm obs}=\int_0^{2R'} F[t_{\rm obs}\delta_D/(1+z)-(\ell/c)] \mathcal{W}(\ell)d\ell$,
 where $\mathcal{W}(\ell)\equiv(3/R')\left[(\ell/2R') - (\ell/2R')^2\right]$ is the geometrical weight function for a homogeneous spherical blob \citep{Zacharias2013ApJ},
 $z$ and $\delta_D$ denote the redshift and the Doppler factor of the source, respectively.
 Its effect on the resulting LCs is to delay the onset of the flare rise and to smooth out any sharp features at the light-crossing timescale.

The time lags between two different LCs are calculated using three different methods: the discrete Fourier transform (DFT) method, the normalized flux matching delay (NFMD) method, and the cross-correlation function (CCF) method. 
These methods are briefly described in the following subsection.

\subsection{Method of time lags}

In the present work, an individual flaring event is generated by introducing an impulsive electron injection with a duration of $t'_{\rm inj}=2R'/c$, represented by a step function: $f'_{\rm e}(t')=1$ for $t' \in[0, 2R'/c]$, otherwise $f'_{\rm e}(t')=0$. 
The Fourier lag-frequency spectra between two LCs at different bands are then calculated numerically using the standard DFT method. 
In the Fourier frequency domain ($\nu_f$), the time lags are obtained by dividing the phase lags $\Delta\phi$ by $2\pi \nu_f$, where $\Delta\phi$ is derived from the argument of the cross-spectrum between a pair of LCs at two energy channels.
After neglecting the nonlinear effects of time-dependent SSC cooling, this approach exhibits a key advantage: the phase lags are independent of the temporal evolution pattern of injection rate fluctuations \citep{Hu2024ApJ}.
The lag-frequency spectra are extremely helpful for probing the intrinsic time lags associated with nonlinear SSC cooling.

To intuitively understand the results, we employ the simple NFMD method to demonstrate the temporal evolution of the delays induced by the model presented here.
The evolution of the time delays is evaluated by comparing the time difference between two LCs when they reach the same normalized flux, and by computing this time difference throughout the flare \citep{Perennes2020A&A}. 
Specifically, we first obtain the normalized fluxes of the hard ($F_{\rm hard}$) and soft ($F_{\rm soft}$) X-ray LCs, and then identify two time points
$t_{\rm hard}$ (for $F_{\rm hard}$) and $t_{\rm soft}$ (for $F_{\rm soft}$) that satisfy the equality $F_{\rm hard}(t_{\rm hard})=F_{\rm soft}(t_{\rm soft})$. 
The instantaneous time delay between the two LCs is thus defined as $\tau\equiv t_{\rm hard}-t_{\rm soft}$.

Alternatively, the lags between different energy bands are estimated using the widely adopted CCF method \citep{Peterson1998PASP}, which uses the full LCs to evaluate time delays.
The measured time delay is quantified using the CCF centroid $\tau_{\rm cent}$, which is built based on all points with correlation coefficients in excess of 80\% of the CCF peak value (i.e., the maximum correlation coefficient). 
The primary reason is that $\tau_{\rm cent}$ can account for the CCF asymmetries that are usually seen in AGN variability.
 In particular, the lags measured with peak and centroid values are consistent with each other when the CCFs are smooth and symmetric functions.

Here, a soft/negative lag indicates that the low-energy variations lag behind the high-energy ones. Conversely, a hard/positive lag is defined as the high-energy variations lagging behind the low-energy ones.
Without loss of generality, the two LCs chosen for the comparison are integrated over the energy ranges $0.5-2$ keV and $2-10$ keV.

     \begin{figure}
   \centering
     \includegraphics[width=0.45\textwidth]{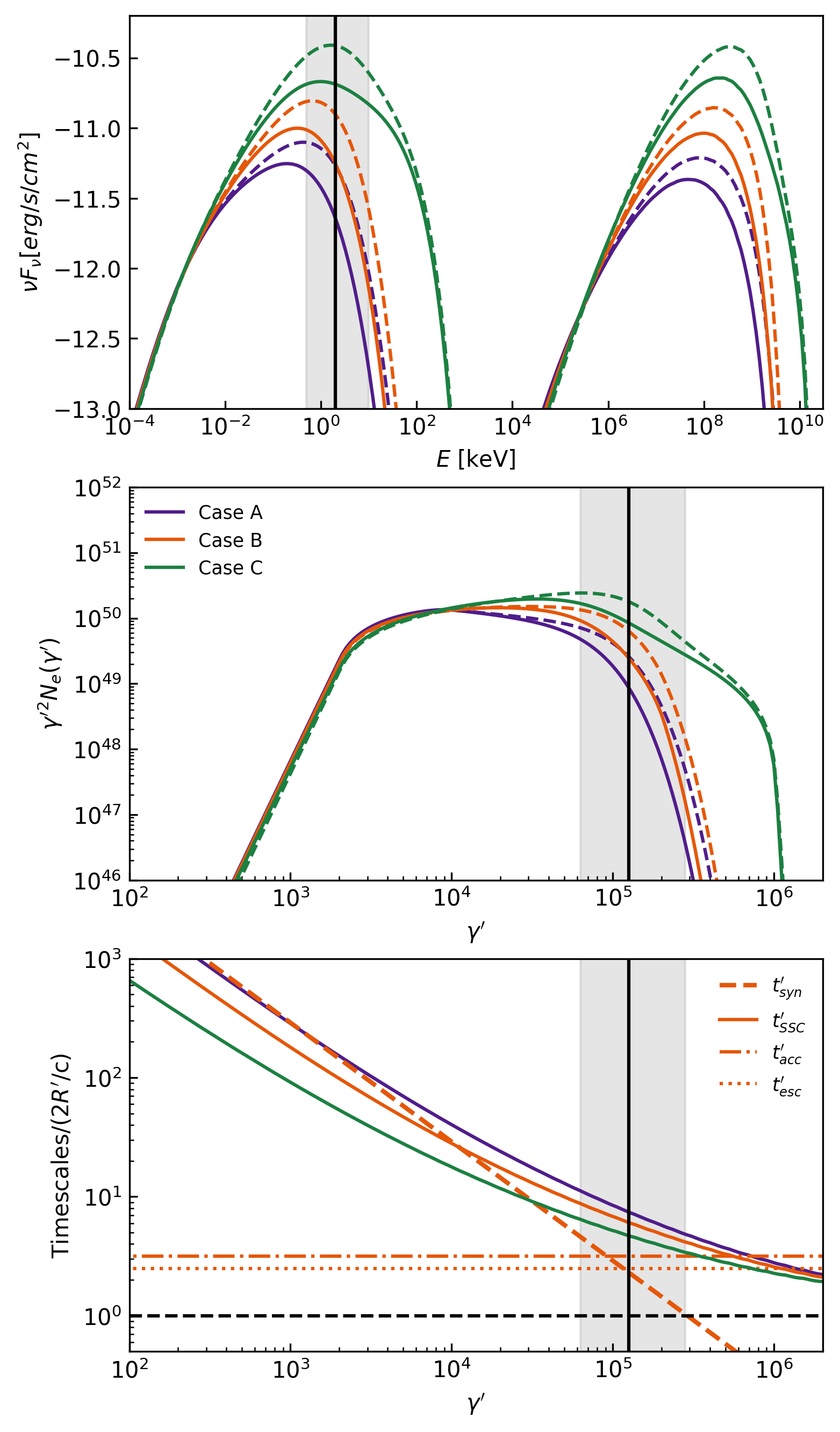}
   \caption{Steady-state SEDs (top), EEDs (middle) and timescales of radiating electrons (bottom) resulting from the three sets of base parameters listed in Table \ref{tabs:basepara},
   illustrating three types of time lags measured in blazar flares.
   The solid lines denote the results calculated by considering nolinear SSC cooling effects, while the dashed lines represent the results accounting only for synchrotron loss.
  The shaded areas in the middle and bottom panels indicate the Lorentz factor ranges of the electrons radiating in the $0.5-10$ keV X-rays, indicated by the shaded area in the top panel.
In the bottom panel, the various timescales in the jet co-moving frame are shown as indicated in the legend, and are plotted in units of $2R'/c$, which is denoted by a horizontal dashed line. 
}
              \label{figs:results}%
    \end{figure}

\begin{table}
	\centering
	\caption{Model parameters for three different lags cases.}
	\label{tabs:basepara}
	\begin{tabular}{lccc} 
		\hline\hline
		Parameters		& Case A	& Case B	&	Case C\\
		\hline
		 $\delta_{\rm D}$	&   40	 	& --		& --\\
         $R' (\rm cm)$ 			&  $10^{15}$	& --		& --\\
		 $B'$ (G)             	&   0.2		& --		& --\\
    $D_0 (\rm s^{-1})$     		& $1.2\times10^{-6} $                 & --      	& --\\
	     $\eta'_{\rm esc}$          & 1.26		& --		& --\\
     $Q_0' ( \rm s^{-1})$   		& $2.69\times10^{46}$		& $2.43\times10^{46}$	&  $5.49\times10^{44}$ \\
    $\gamma'_{\rm i,min} $         & $2\times10^3$     		& --		&  --  \\
    $\gamma'_{\rm i,max}$        &$10^4$		& $2.1\times10^5$		& $10^6$ \\
	  $p$ 			        &  2.45              & 2.45     	& 2.0 \\
		\hline\hline
	\end{tabular}
\end{table}

\subsection{Model  parameter setup}

The model is defined by nine physical quantities (i.e., $B'$, $\delta_{\rm D}$, $R'$, $D_0$, $\eta_{\rm esc}$, $q_e'$, $\gamma_{\rm i,min}'$,  $\gamma_{\rm i,max}'$ and $p$).
In the following analysis, we fix the blob radius $R'=10^{15}$ cm and set $\Gamma=\delta_{\rm D}=40$. The choice of these parameters is intended to produce the observed rapid variability and overcome the photon opacity barrier.
For the magnetic field strength and momentum diffusion coefficient, we adopt typical values consistent with previous works: $B'=0.2$ G and $D_0\simeq1.2\times10^{-6} \rm s^{-1}$ \citep[e.g.,][]{Lewis2016,Hu2021MNRAS,Hu2023ApJ}.
The competition between acceleration and synchrotron cooling then establishes an equilibrium at an energy corresponding to $\gamma'_{\rm eq}\simeq10^5$, 
which corresponds to the observed X-ray emission at  $E_{\rm eq,sy}\simeq1$ keV.
Additionally, we adopt an escape efficiency of $\eta_{\rm esc}=1.26$ \citep{Hu2023ApJ}.
For the injected electron spectra, we specify the minimum Lorentz factor of the injected electrons as $\gamma'_{\rm i,min} = 2\times10^3$, which does not significantly affect the key results of our study.
We adjust the injection rate to ensure that nonlinear SSC cooling plays a non-negligible role,
while the maximum Lorentz factor $\gamma'_{\rm i,max}$ and spectral index $p$ are adjusted to produce three representative cases: a hard/positive lag,
a soft/negative lag, and a transition between these two cases.
Thus, the key difference between the current simulations lies in the injected electron distribution.
The time-dependent emission from the plasma blob is driven by the acceleration process when $\gamma'_{\rm i,max}$ is significantly less than the equilibrium energy $\gamma'_{\rm eq}$.
In contrast, the cooling process dominates variability when $\gamma'_{\rm i,max}$ is much larger than $\gamma'_{\rm eq}$, accompanied by a relatively harder spectral index $p=2$.
For the transition state, $\gamma'_{\rm i,max}$ and $\gamma'_{\rm eq}$ are comparable. 
The basic values of all parameters are reported in Table \ref{tabs:basepara}.

The FP equation is solved numerically via an implicit Crank-Nicholson(CN) scheme, which offers the advantage of unconditional stability. 
To ensure the evolution is fully resolved numerically, we use a time step of $\Delta t'=t'_{\rm lc}/500$ ($t'_{\rm lc}\equiv2R'/c$ is the light-crossing time) and a 300-point energy grid over the range $1 \leq \gamma' \leq 10^7$.

  \begin{figure*}
   \centering
  \includegraphics[width=0.95\textwidth]{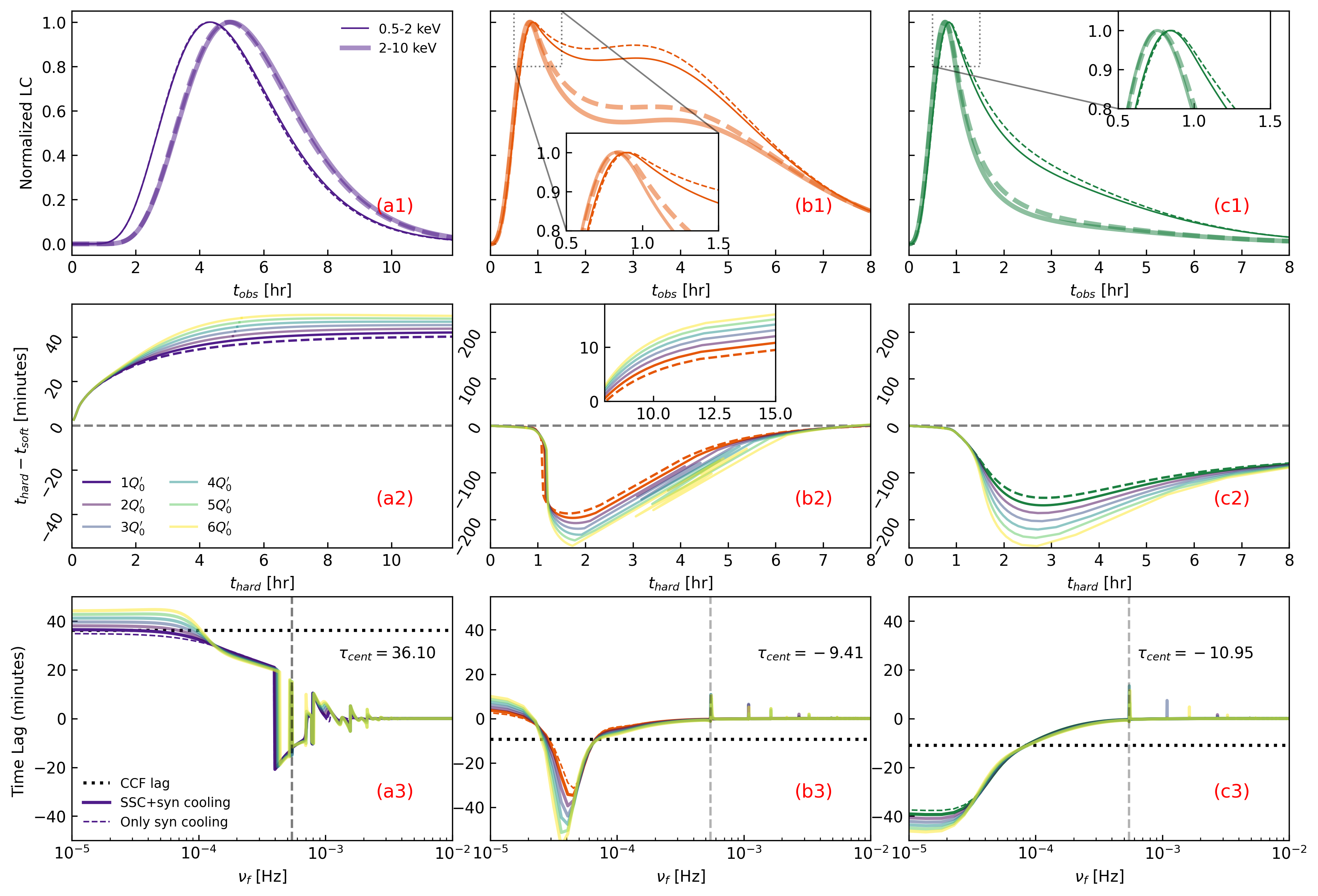}
   \caption{Normalized LCs, instantaneous time delay and Fourier lag-frequency spectra measured between the hard and soft bands for the three representative lag regimes.
The normalized LCs for the three cases are shown in the top panels, where the thick and thin lines represent the hard and soft LCs, respectively.
The resulting Fourier and CCF time lags are shown in the bottom panels. 
 The dashed lines show the results without considering nonlinear SSC cooling, while the solid lines represent the results including SSC cooling.
The gradient-coloured lines show the results calculated for different injection rates from $Q'_0$ to $6Q'_0$.
The corresponding instantaneous lags are presented in the middle panels.
}
              \label{figs:flare}%
    \end{figure*}

\section{Results}\label{sec:result}

In Fig. \ref{figs:results}, we show the steady-state spectral energy distributions (SEDs), electron energy distributions (EEDs), and characteristic timescales resulting from three representative model parameter sets under continuous electron injection.
To illustrate the effect of SSC cooling, we display these results both with and without SSC cooling in the figure.
Obviously, the synchrotron and SSC radiation are comparable, and the fluxes and positions of the two peaks obtained when neglecting SSC cooling
are higher than those derived with consideration of both synchrotron and SSC cooling processes. 
These results verify that nonlinear SSC cooling for our adopted parameter sets plays a non-negligible role in producing the X-rays, though the electrons responsible for X-rays are dominated by linear synchrotron cooling losses, 
due to the decrease in scattering efficiency in the KN regime.
To allow a direct comparison between the numerical resolution and the shortest physical timescales,  
all timescales displayed in the bottom panel of the figure are normalized to $t'_{lc}$.
It is clear that $t'_{lc}$ is comparable to the minimum value of the cooling, acceleration, and escape timescales across the entire relevant electron energy range.
This guarantees that the competition between these processes is accurately captured in the solution of the time-dependent FP equation.

For the three representative situations, the normalized soft and hard X-ray LCs are displayed in the top panels of Fig. \ref{figs:flare},
while the evolution of EEDs is provided in Appendix \ref{sec:append1}.
The instantaneous time delays are shown in the middle panels of the figure,
and the bottom panels present the Fourier lag-frequency spectra and classical CCF lags derived from these LCs.
Additionally, it is necessary to highlight the impact of SSC cooling on the time lags.
In the middle and bottom panels of Fig. \ref{figs:flare}, we show the instantaneous time delays and Fourier frequency-resolved lags resulting from varying the injection rate,
which determines the strength of the nonlinear SSC cooling.
We perform this study by continuously increasing the electron injection rate from $Q_0'$ to $6Q_0'$.

It can be seen that the profiles of the X-ray LCs exhibit distinct behaviors across the three representative situations.
In Case A (hard lag), the X-ray LCs show an approximately symmetric profile, whereas Cases B and C (soft lag) produce asymmetric X-ray LCs.
A prominent bump is observed in the declining phase of the LC generated in Case B.

On the other hand, the Fourier frequency-resolved lags also display different characteristics below the frequency $t_{\rm lc}^{-1}$, and their spectral shapes are similar to those of the instantaneous lags shown in the middle panels of the figure.
For Cases A and C, the time lags in the low-frequency regime of the lag-frequency spectra remain approximately constant.
Compared to Case C, the lag-frequency spectrum in Case A features a break from positive to negative lags at frequency $\nu_f\sim4\times10^{-4}$ Hz, followed by a gradual return to near-zero lags at higher frequencies.
Moreover, the lags may oscillate in the high-frequency regime, but precise quantification is limited by the nonlinear nature of Fourier lag calculations.
For Case B, the lag-frequency spectrum shows small positive values at low frequencies (large times), which then transitions to negative values at high frequencies (small times), and finally returns to zero lags at higher frequencies.

A comparison of the variation behaviors of lags obtained with the two different methods 
 confirms that the time delays measured in all three cases vary significantly throughout the flare evolution.

Moreover, we note that the low-frequency Fourier time lags in Case A are in good agreement with the CCF lag, which is very reliable for symmetric LCs.
For the asymmetric LCs in Case C, the measured CCF lag is significantly smaller than the low-frequency Fourier time lags
and is comparable to the CCF lag measured in Case B. This implies that it is difficult to distinguish the time delays in Cases B and C.
This is consistent with the fact that CCF lags, evaluated using the full LCs, average out frequency-dependent lag variations and are less sensitive to transitional behavior compared to Fourier methods.

\begin{figure}
   \centering
  \includegraphics[width=0.45\textwidth]{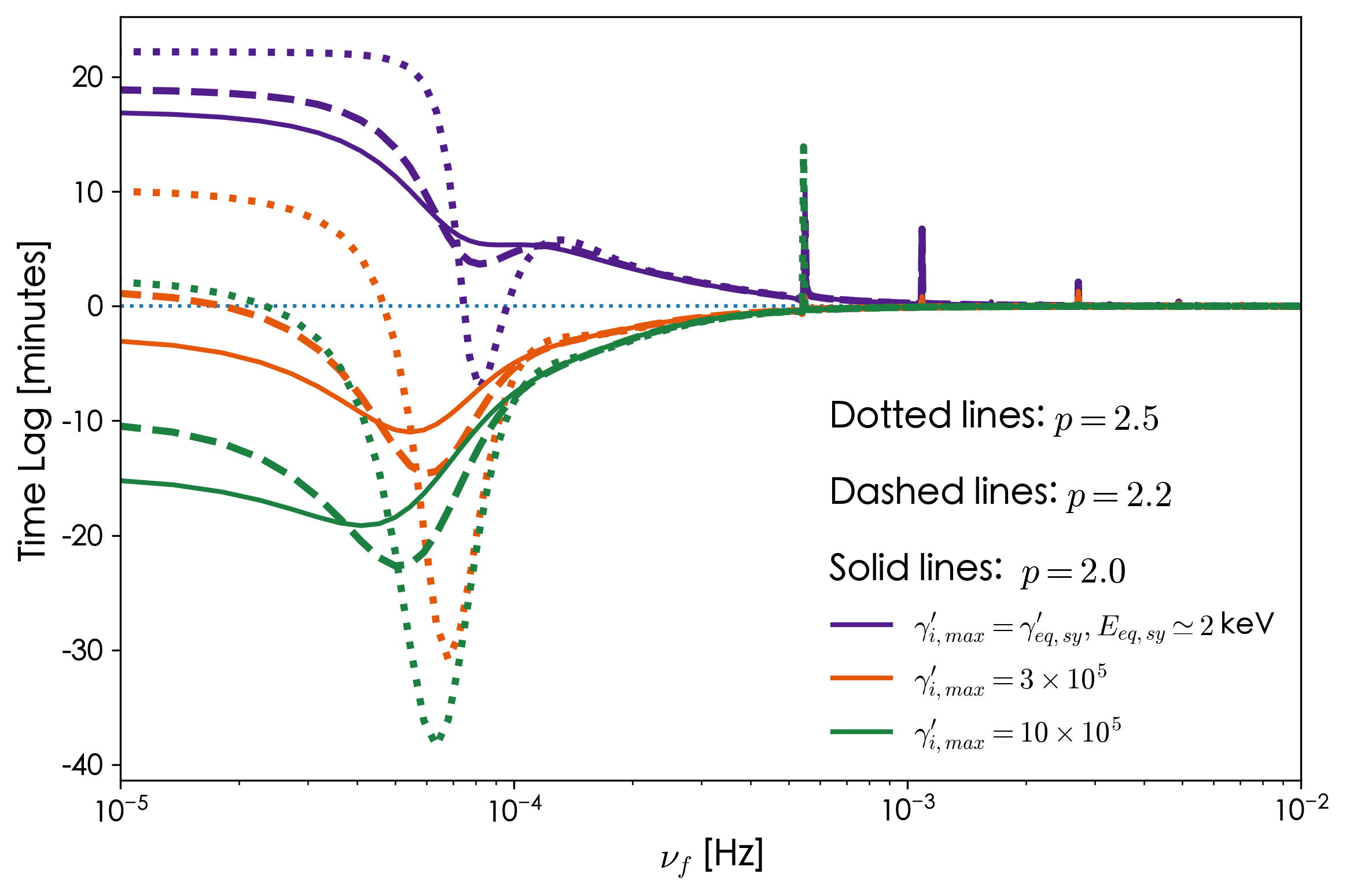}
   \caption{The influence of maximum Lorentz factor $\gamma'_{\rm i,max}$ and the electron spectral index $p$ on the Fourier lag-frequency spectra.
    The results calculated with three values of $\gamma'_{\rm i,max}$ are labelled in the figure legend. For each $\gamma'_{\rm i,max}$, the solid, dashed and dotted lines correspond to $p=2.0, 2.2$ and 2.5, respectively.
}
              \label{figs:diss}%
    \end{figure}

We stress that the negative time lags arise from the radiative cooling of the highest-energy electrons. 
Figure \ref{figs:diss} displays the effects of the maximum Lorentz factor $\gamma'_{\rm i,max}$ and power-law index $p$ on the Fourier lag-frequency spectra, calculated for $E_{\rm eq,sy}\simeq2$ keV.
 As $\gamma'_{\rm i,max}$ decreases to $\gamma'_{\rm eq,sy}$, the Fourier time lags at low frequencies gradually shift from negative to positive values
while a steeper spectral index $p$ leads to a sharper transition from positive to negative lags. 

Additionally, a direct comparison between the lag-frequency spectra obtained with and without incorporating nonlinear SSC cooling
demonstrates that SSC cooling can enlarge the time delays across the three situations.
The amplifying effect is predominantly attributed to the suppression of radiative cooling by STA.

In the following subsections, we present a more detailed analysis of the effects of STA and nonlinear SSC cooling on the Fourier lag-frequency spectra.

    \begin{figure}
   \centering
  \includegraphics[width=0.45\textwidth]{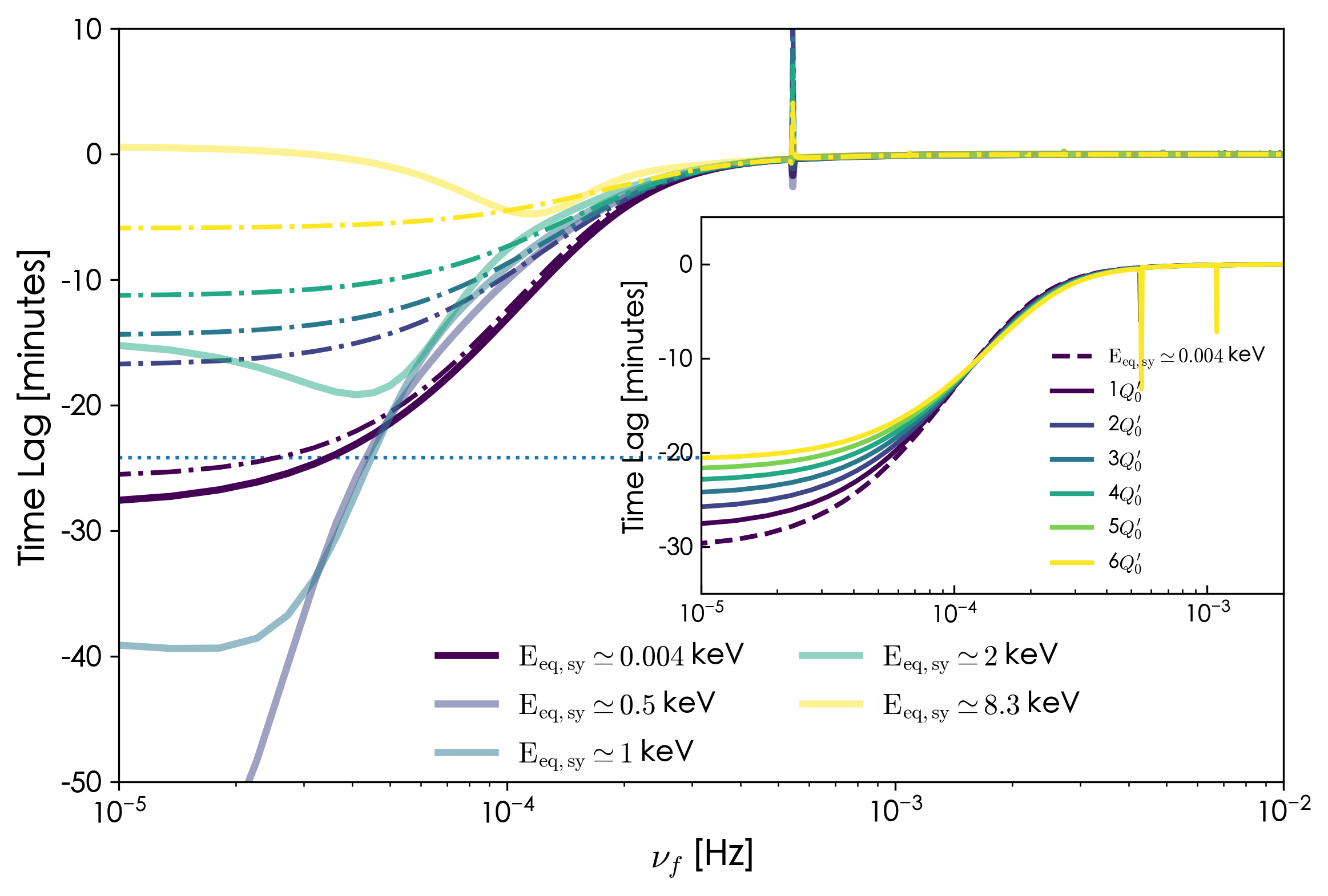}
   \caption{The influence of the diffusion coefficient $D_0$ on the Fourier lag-frequency spectra.
The gradient-coloured solid lines display the results calculated with different values of $E_{\rm eq,sy}$ varying from 0.004 to 8.3 keV,
 while the corresponding results from the pure cooling model are denoted by dash-dotted lines.
In the inset, the gradient-coloured lines show the results calculated with $E_{\rm eq,sy}=0.004$ keV for different injection rate from $Q'_0$ to $6Q_0'$,
and the dashed line denotes the result without accounting for SSC cooling.
}
              \label{figs:diss2}%
    \end{figure}

\subsection{Effects of STA on the Fourier time lags}
In addition to Case A, STA plays an important role in determining the lag-frequency spectra for Cases B and C.
To clarify the point, we conduct simulations where the equilibrium synchrotron energies are $E_{\rm eq,sy}\simeq 0.004, 0.5, 1, 2,$ and 8.3 keV, 
corresponding to the diffusion coefficients $D_0\simeq7.8\times10^{-8}, 8.6\times10^{-7}, 1.2\times10^{-6}, 1.75\times10^{-6}$ and $3.5\times10^{-6}\ s^{-1}$, respectively.
Meanwhile, we also calculate the lag-frequency spectra resulting from the pure cooling model without accounting for the STA,
where we adopt the same values for the other physical parameters as those used in the above simulations. 
For the pure cooling model, the main difference is the electron escape timescale, with values of $t'_{\rm esc}\simeq 76, 7, 5, 3.4,$ and $1.7R'/c$.
All these results are shown in Fig. \ref{figs:diss2}.

These results indicate that the time delays can be either too small to be detectable or large, up to an hour, depending on $D_0$.
It can be seen that the amplitudes of the lags measured in the pure cooling model decrease with decreasing $t_{\rm esc}'$.
This behavior is consistent with the theoretical prediction that electrons lack sufficient time to undergo radiative cooling before escaping the system when $t_{\rm esc}'$ is shorter than the radiative cooling timescale.
Conversely, as $t_{\rm esc}'$ increases, a greater fraction of electrons will experience cooling before escaping.
In particular, the electrons responsible for the X-rays are in the fast cooling regime when $t'_{\rm esc}\simeq 76 R'/c$.
The lag magnitudes reach a maximum value and are in agreement with the analytical estimation of the lags resulting from pure synchrotron cooling, indicated by the dotted line in Fig. \ref{figs:diss2}.

On the other hand, the lag magnitudes resulting from the re-acceleration model are larger than those obtained in the pure cooling model for all cases except when $E_{\rm eq,sy}\simeq 8.3$ keV.
This behavior can be well understood because the radiative cooling-driven transfer of electrons from high to low energies is suppressed by the STA process, which continuously accelerates low-energy electrons to higher energies.
Consequently, STA plays a role in amplifying the lag magnitude when $E_{\rm eq,sy} < 8.3$ keV or $D_0<3.5\times10^{-6}\ s^{-1}$.
Moreover, our extended numerical simulations indicate that the amplifying effect of the STA process reaches a maximum level when $E_{\rm eq,sy}$ is close to 0.5 keV,
whereas it diminishes as $E_{\rm eq,sy}> 0.5$ keV.
This is naturally expected, as when $E_{\rm eq,sy}> 0.5$ keV, STA dominates over radiative cooling for the electrons that were previously driven by radiative cooling when $E_{\rm eq,sy}< 0.5$ keV. 
In other words, the weakening of the amplifying effect essentially reflects a gradual reduction in the fraction of cooling-driven electrons relative to acceleration-driven ones (please see Section~\ref{subsec:secB} for further discussion).

For these reasons, the lag magnitudes in the re-acceleration model exhibit a non-monotonic variation: they increase as $D_{0}$ increases for $E_{\rm eq,sy}\lesssim 0.5$ keV,
and then decrease for $E_{\rm eq,sy}> 0.5$ keV.
In particular, a nearly zero time lag is observed when $E_{\rm eq,sy}\simeq 8.3$ keV, indicating that a dynamic equilibrium is established between the electrons responsible for the soft and hard X-ray emission. 


    \begin{figure}
   \centering
  \includegraphics[width=0.45\textwidth]{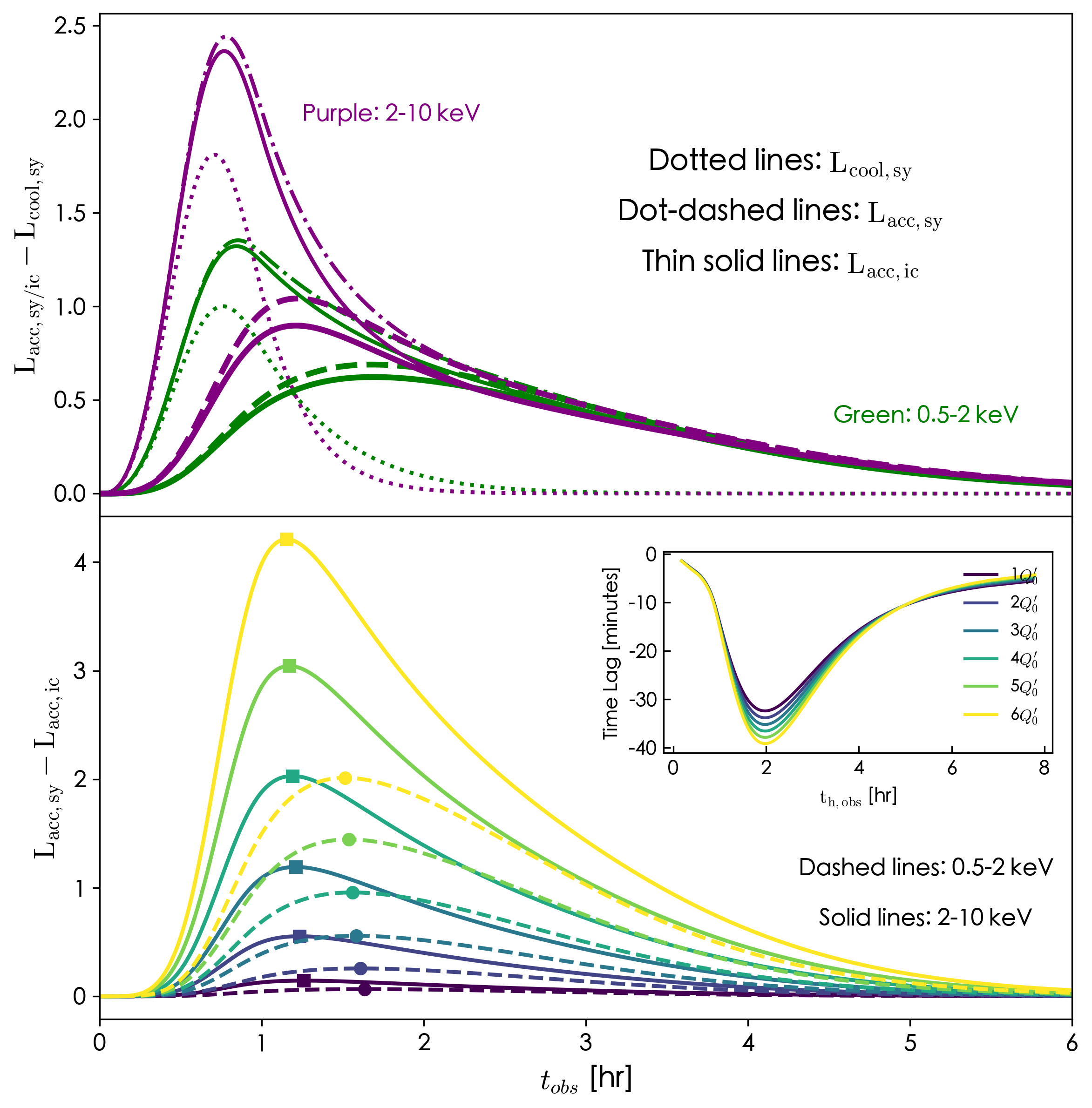}
   \caption{The effects of STA suppression and SSC cooling on the time lags between soft ($0.5-2$ keV) and hard ($2-10$ keV) X-rays for $E_{\rm eq,sy}\simeq2$ keV.
   In the upper panel, thick dashed and solid lines denote $L_{\rm acc,sy}-L_{\rm cool,sy}$ and $L_{\rm acc,ic}-L_{\rm cool,sy}$, respectively (see text for detail).
}
              \label{figs:lc}%
    \end{figure}
    
\subsection{Effects of SSC cooling on the Fourier time lags}\label{subsec:secB}

Our results indicate that SSC cooling can enhance the time delays across the three situations when STA plays an important role in reshaping the X-ray spectra.
In this subsection, we focus on clarifying this point.

For Case A, the increase in the time lags arises naturally from heightened energy loss induced by the SSC process.
This is because the an increase in radiative cooling results in a longer time required for low-energy electrons to reach higher energies.

For Case C, the effect of SSC cooling on the Fourier lag-frequency spectra differs between the regimes below and above $E_{\rm eq,sy}\sim 0.5$  keV.
For $E_{\rm eq,sy}<0.5$ keV, radiative cooling fully dominates over STA for the electrons responsible for the 0.5–10 keV emission.
Physically, we predict that the lag magnitude decreases as SSC cooling strength increases (see the results in the inset of Fig. \ref{figs:diss2}), which is caused by an increase in the injection rate.

However, this trend is opposite to that observed for $E_{\rm eq,sy}>0.5$ keV.
Under this condition, the lag magnitudes increase with the strength of SSC cooling (also see Fig. \ref{figs:flare}). 
To better clarify this phenomenon, we extract the radiation component associated with the electrons arising from the competition between the acceleration of low-energy electrons and the radiative cooling of high-energy electrons.
To achieve this goal, the soft and hard X-rays generated in the pure cooling model are subtracted from the X-rays produced by the re-acceleration model.
This approach effectively eliminates the contribution from injected electrons.
For example, the results calculated for $E_{\rm eq,sy}\simeq2$ keV are shown in Fig. \ref{figs:lc}.
In the upper panel of the figure, we show the differences $L_{\rm acc,sy/ic}-L_{\rm cool,sy}$,
where $L_{\rm acc,sy/ic}$ denotes the LCs produced by the re-acceleration model without and with incorporating nonlinear SSC cooling, respectively,
and $L_{\rm cool,sy}$ denotes the LCs produced in the pure synchrotron cooling model. 
In the bottom panel, we display the differences $L_{\rm acc,ic}-L_{\rm acc,sy}$, representing the contribution from SSC cooling.
These results are calculated for injection rates increasing from $Q_0'$ to $6Q'_0$. 
It is clearly seen that the contribution from SSC cooling increases with the injection rate, and the decline phase of soft X-ray emission is slower than that of hard X-ray emission. 
The time lags between the soft and hard X-ray bands increase with the injection rate, 
with maximum lags occurring at $\sim 2$ hr and instantaneous time delays presented in the panel inset.

As observed in the figure, due to the STA suppression effect, both soft and hard X-rays obtained after subtracting the synchrotron cooling component peak significantly later than those in the pure cooling model,
and the resulting time lags between the two bands are substantially amplified in the re-acceleration model. 
Although the SSC process enhances electron radiative cooling, which in turn reduces the maximum electron energy attainable via STA,   
the electrons emitting soft X-rays can be slowly accelerated via STA, which remains dominant over radiative cooling for the electrons producing low-energy X-rays.
Thus, the electrons emitting soft X-rays can be compensated, and therefore the variation in soft X-rays is more moderate compared with that in hard X-rays. 
As a result, the lag magnitudes increase with the injection rate.

\begin{figure}
   \centering
    \includegraphics[width=0.45\textwidth]{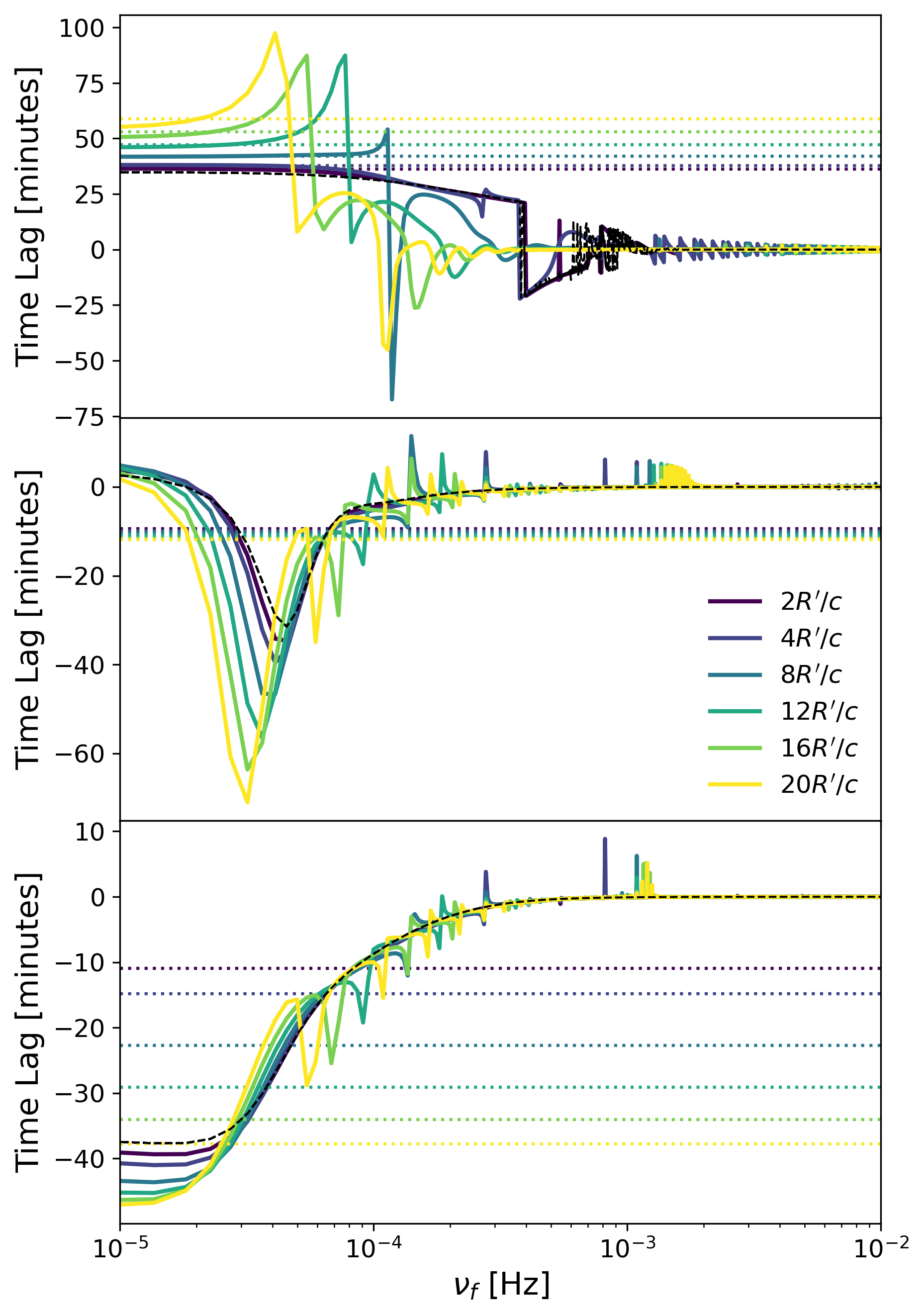}
   \caption{
Fourier lag-frequency spectra calculated with different injection time-scales varying from $2R'/c$ to $20R'/c$.
Top, middle and bottom panels show the results for Case A, B, and C, respectively.
The black dashed lines denote the results without incorporating nonlinear SSC cooling.
The injection time-scales are labelled at the right lower corner of the middle panel in the figure.
        }
   \label{figs:tlc}%
    \end{figure}

Additionally, we expect a positive correlation between the lags and flare durations based on the knowledge of SSC cooling effects.  
The primary reason is that the strength of SSC cooling is positively related to the number of electrons in the system. 
For the purpose of illustration, the influence of injection time on the Fourier lag-frequency spectra is presented in Fig. \ref{figs:tlc},
where we also show the Fourier lag-frequency spectra calculated without incorporating nonlinear SSC cooling, which are independent of the injection profile.
To further verify the reliability of this correlation, we also report the time lags measured via the CCF method, denoted by the dotted lines in the figure.
In the simulations, we choose six total injection times, increasing from $2R'/c$ to $20R'/c$.  
As shown in Fig. \ref{figs:tlc}, the time lags at low frequencies appear to be correlated with the injection time, although the high-frequency lags are distorted by nonlinear SSC cooling effects.
Moreover, this variation trend is in consistent with that of the CCF lags.

\section{Discussion and Summary}\label{sec:discu}

 In this work, we focus on an interesting acceleration scenario in which a population of pre-accelerated electrons with a power-law distribution is injected into the emission region
and then further accelerated via STA. 
We have numerically solved the kinetic equation for relativistic electrons
to systematically investigate the X-ray Fourier time lags, which serve as a powerful diagnostic tool for probing jet physics in HBLs. 
Compared to previous theoretical studies, our work extends the analytical models of \cite{Finke2014ApJ,Finke2015ApJ} and \cite{Lewis2016} by incorporating nonlinear SSC cooling,
which plays an important role in determining the LCs and energy spectra observed in blazars. 
In the following, we conclude with a summary of our main results, a discussion of implications for interpreting observations, and our future research plans.

Compared to previous studies, the re-acceleration model reveals three distinct time-lag regimes: hard/positive lags, soft/negative lags, and a transitional regime between them. 
Thus, it can provide a unifying framework for interpreting the diverse time-lag signatures detected in HBLs \citep[e.g.,][]{Zhongli2019ApJ,Zhongli2021ApJ,Noel2022ApJS,Devanand2022ApJ,Das2023MNRAS,Gokus2024MNRAS}, 
in which the X-ray spectra exhibit a log-parabolic shape imprinted by the STA process.
These time lags carry rich information about energy transport, particle acceleration, and radiative cooling in the most energetic jets, 
thereby imposing stringent constraints on their extreme physical conditions.

Of particular interest is the existence of a transition between positive and negative lag regimes. 
In this transition, the lag-frequency spectrum is characterized by a smooth transition from positive values at low frequencies to negative values at high frequencies, 
reflecting the coexistence of acceleration-dominated low-energy electrons and cooling-dominated high-energy electrons. 
Therefore, this transition may be considered as evidence of the re-acceleration of relativistic electrons in the system.
Moreover, we note that the lag measured by the CCF method can average out frequency-dependent lag variations and therefore should be less sensitive to transitional behavior compared to Fourier methods.
The weak or undetectable lags measured via a time-domain approach in some flares may hint at the presence of the transition, although the nearly zero lags predicted in the model may
indicate a dynamic equilibrium between the electrons responsible for the soft and hard X-rays.

We find a significant difference in the time-lag signatures due to the presence or absence of the STA process.
In particular, inclusion of STA results in amplification and modification of time lags in the transitional and cooling-dominated regimes. 
The magnitude of time lags with incorporating the STA process is larger than that in the pure cooling model, 
as STA suppresses radiative cooling-driven electron energy loss by continuously reaccelerating low-energy electrons. 
Since soft X-rays are slowly sustained by reaccelerated lower-energy electrons while hard X-rays are more sensitive to additional SSC cooling,
the resulting time lags for $E_{\rm eq,sy}>0.5$ keV can be further enlarged by SSC cooling.
Thus, relatively large lags may be expected to be observed in TeV-bright flares \citep[e.g.,][]{Abramowski2012A&A, Aleksi2015A&A, Abeysekara2017ApJ,Abeysekara2020,Acciari2020ApJS,MAGIC2021A&A}, 
where SSC emission should be prominent. 
Additionally, SSC cooling effects provide a natural explanation for the trend between lag amplitude and flare duration: the larger the flare duration, the larger the lag \citep{Zhang2002MNRAS,Brinkmann2003A&A}.

Besides, the Fourier lag--frequency spectrum is helpful for probing the underlying energetic processes powering rapid X-ray variability.
In particular, Mrk 421 and PKS 2155-304 are two of the brightest and most studied HBLs in the X-ray band.
Many studies have demonstrated that their X-ray spectra are best described by a log-parabolic model, supporting the idea that STA plays an important role in reshaping the X-ray spectra \citep[e.g.,][]{Massaro2008A&A,Tramacere2007,Tramacere2009A&A}.
\cite{Hu2024ApJ} performed Fourier analysis of 10 archived \XMM observations of Mrk 421, each with an exposure time exceeding 40 ks.
The results reveal that, for most observations, lags predominantly appear in the low Fourier-frequency bins where color noise dominates,
and the magnitude of time lags increases with the energy difference $\Delta E$ between two different LCs in X-rays.
These results are consistent with the predictions of standard one-zone SSC models and naturally arise from the temporal evolution of the emitting electron spectra. 
Later, using three archived \XMM observations of PKS 2155-304, each with a total exposure of $>80$ ks,
\cite{Hu2025A&A} found that PKS 2155-304 exhibits the same $\Delta E-lag$ behavior reported for Mrk 421,
and the lag-frequency spectra measured in ObsID 0124930601 are consistent with characteristics of the transitional regime predicted by the model.
Moreover, the authors demonstrated that the observed X-ray LCs agree with those from theoretical simulations by varying the injection rate of electrons, 
and suggested that the energetic electrons should be injected by shocks formed in a weakly magnetized jet.

Very recently, \cite{MAGIC2025arXiv250908686M} demonstrated that the majority of sub-hour flux and spectral variations in Mrk 421 during substantial flaring activity in April 2013 
 can be explained by changes in the luminosity and slope of the injected electron spectra.
 The team claimed that the injected electrons may be associated with a shock embedded in the jet plasma. 
This result supports our preliminary conclusion inferred from combining lag-frequency spectra with SED modeling in \cite{Hu2024ApJ} and \cite{Hu2025A&A}, 
and confirms the feasibility of our approach and the reliability of our findings.  

 \begin{figure}
   \centering
    \includegraphics[width=0.45\textwidth]{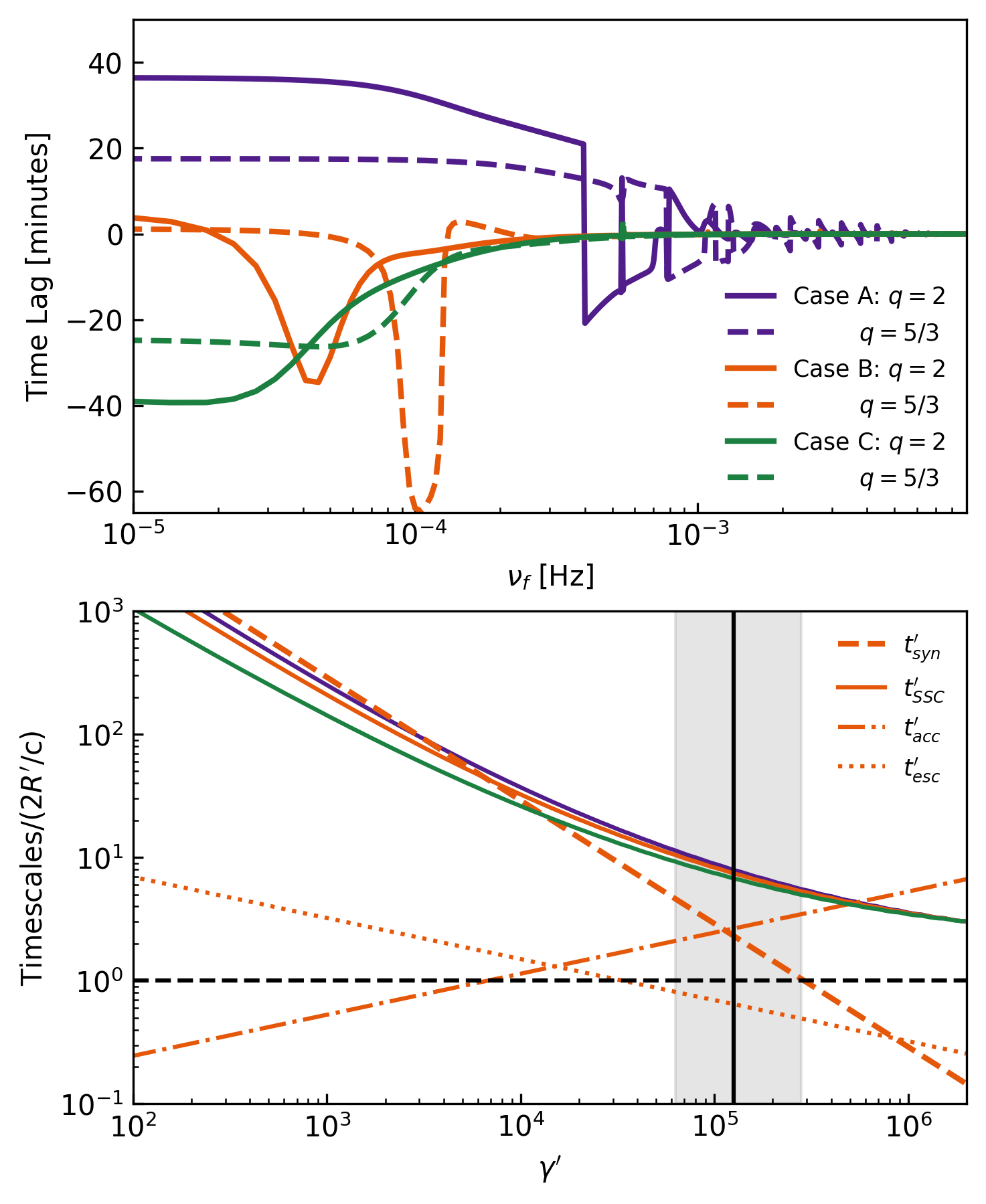}
   \caption{
   Upper panel compares the Fourier lag–frequency spectra obtained for the turbulence indices $q=5/3$ (Kolmogorov) and $q=2$ (hard-sphere limit).
   Lower panel presents the characteristic timescales of particle acceleration, radiative cooling, and escape for $q=5/3$.
        }
   \label{figs:q5/3}%
    \end{figure}

To further examine its capability to constrain the physical nature of stochastic acceleration in HBL jets,
we repeated the key simulations and computed the corresponding Fourier lag–frequency spectra for the Kolmogorov turbulence index $q=5/3$.
In these calculations, we used $D_0\simeq7.8\times10^{-5}~s^{-1}$ and set $t'_{\rm esc}(\gamma')t'_{\rm acc}(\gamma')=(27/11)(R/\beta_{\rm A}c)^2$ 
\footnote{The relation can be inferred from the connection between the energy-diffusion coefficient and the spatial diffusion coefficient $D_\gamma \kappa_\gamma=(\beta_Ac\gamma/3)^2$ \citep{Stawarz2008,Tramacere2011ApJ}, where $\beta_{\rm A}$ is the Alfv\'{e}n velocity in units of $c$.}
with $\beta_{\rm A}=0.6$, while keeping all other parameters unchanged.
This setup yields $\gamma'_{\rm eq}\simeq10^5$ (Fig.~\ref{figs:q5/3}) and a soft X-ray spectrum similar to that obtained for $q=2$ (Fig.~\ref{figs:sed_q}).
From Fig.~\ref{figs:q5/3}, we find that the overall features of the lag–frequency spectra remain qualitatively consistent with the hard-sphere case, 
while quantitative differences in the lag amplitude and characteristic frequency arise for different $q$. 
Combining these distinct quantitative signatures with broadband SED modeling may provide a promising approach to distinguish between different turbulence configurations. 

Since both first-order (\emph {Fermi}-I) and second-order (\emph {Fermi}-II) acceleration processes may occur simultaneously in blazar jets, particularly in Mrk 421 \citep {Tramacere2009A&A}, 
we extend our study of STA and SSC cooling to assess whether and how the \emph {Fermi}-I process can modify the lag–frequency spectra, as illustrated in Fig.~\ref {figs:lags_q}.  
Here, we consider the systematic energy gain $A(\gamma')=A_{\rm sh}\gamma'$ caused by the \emph{Fermi}-I process.
Our results show that the effect of the \emph{Fermi}-I mechanism can be negligible within the framework of our present model for Case A, when the \emph{Fermi}-I acceleration timescale satisfies $t'_{\rm sh} > t'_{\rm D}/2$. 
In contrast, the additional \emph{Fermi}-I mechanism can significantly modify the lag-frequency spectra in Cases B and C. 
Nevertheless, our main conclusions should remain unchanged even after incorporating the \emph{Fermi}-I process.

We stress that the one-zone homogeneous model is fundamentally important for understanding the origin of rapid X-ray variability, as well as the geometry of the innermost region in HBLs.
In the present work, a single impulsive electron injection is assumed to mimic a flare, but real jet flares may arise from continuous injection or multiple injection episodes \citep{Roken2009,Roken2018,Hu2021MNRAS,Thiersen2022ApJ,Hu2023ApJ},
or they may also be related to variations in the most energetic electrons $\gamma'_{\rm i,max}$ and/or power-law index $p$ \citep{MAGIC2025arXiv250908686M}.
On the other hand, in reality, blazar jets may exhibit radial variations in physical properties such as magnetic field strength, turbulence intensity, and particle density \citep{Blandford1979, Konigl1981}, 
which could introduce additional time-lag signatures \citep{Perennes2020A&A}. 
Future work should incorporate inhomogeneous jet models to explore the interplay between spatial structure and time lags,
and exploring time lags under different injection histories will help bridge theoretical models with observational constraints, enhancing our ability to interpret the complex variability observed in blazars.

Finally, we anticipate that the upcoming Advanced Telescope for High Energy Astrophysics \footnote{https://www.the-athena-x-ray-observatory.eu/es}({\it Athena}) 
will enable precise measurements of lag-frequency spectra, facilitating direct comparisons with our model predictions. 
This is because {\it Athena} outperforms \XMM across the board in energy resolution, effective area, and time resolution. 
Additionally, operating in a halo orbit at the Sun-Earth Lagrangian point (L1), {\it Athena} avoids Earth occultations and radiation belts, 
enabling continuous observations of up to 40–50 hours—critical for capturing low-frequency variability in Fourier analysis.
Combining precisely measured lag--frequency spectra with SED modeling will be of great potential value for providing insights into the extreme physical conditions in the compact emission regions of relativistic jets,
and will advance our understanding of the physical nature of blazars.

\begin{acknowledgments}
We thank the anonymous referee for constructive comments that helped improve and clarify the paper.
This work is supported by the National Key Research and Development Project of China (grant No.2021YFA0718500).
We acknowledge the support of National Natural Science Foundation of China (NSFC-12263003, -12403048 and -12033006). 
WH acknowledges support from the Natural Science Foundation of Jiangxi Province (20252BAC240161), Jiangxi Provincial Key Laboratory of Modern Agricultural Equipment(Grant No. 20242BCC32127) 
and the PhD Starting Fund program of JingGangShan University under Grant No. 2017KFW001.
\end{acknowledgments}

\appendix

\section{Theoretical LCs and temporal evolution of the electron spectra}\label{sec:append1}

\begin{figure*}
   \centering
  \includegraphics[width=0.9\textwidth]{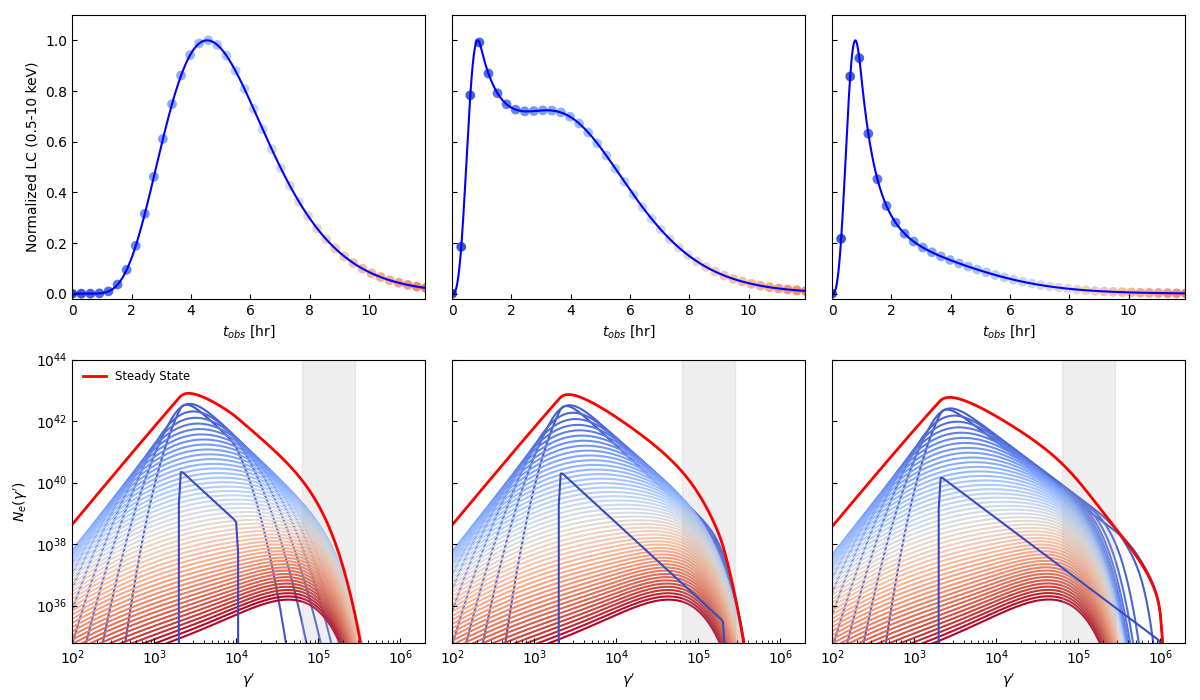}
   \caption{
Theoretical LCs in the $0.5-10$ keV X-ray band (upper panel) and the evolution of the emitted electron spectra (lower panel). 
In the lower panel, the red solid lines show the steady-state electron spectra resulting from constant injection,
and the colour--gradient lines represent the time-dependent electron spectra calculated at the times corresponding the dots in the upper panel.
The shaded areas indicate the range of electron Lorentz factors corresponding to the X-ray emission presented in the upper panels. 
        }
   \label{figs:eedtlc}%
    \end{figure*}

In Fig. \ref{figs:eedtlc}, we show the theoretical LCs in the X-ray band and the time–dependent evolution of the electron energy spectra generated
by applying an impulsive injection with a duration of $t'_{lc}=2R'/c$. The values of the physical parameters are reported in Table \ref{tabs:basepara}.

\section{Steady-state solutions obtained for $q=5/3$}\label{sec:append2}

\begin{figure}
   \centering
  \includegraphics[width=0.45\textwidth]{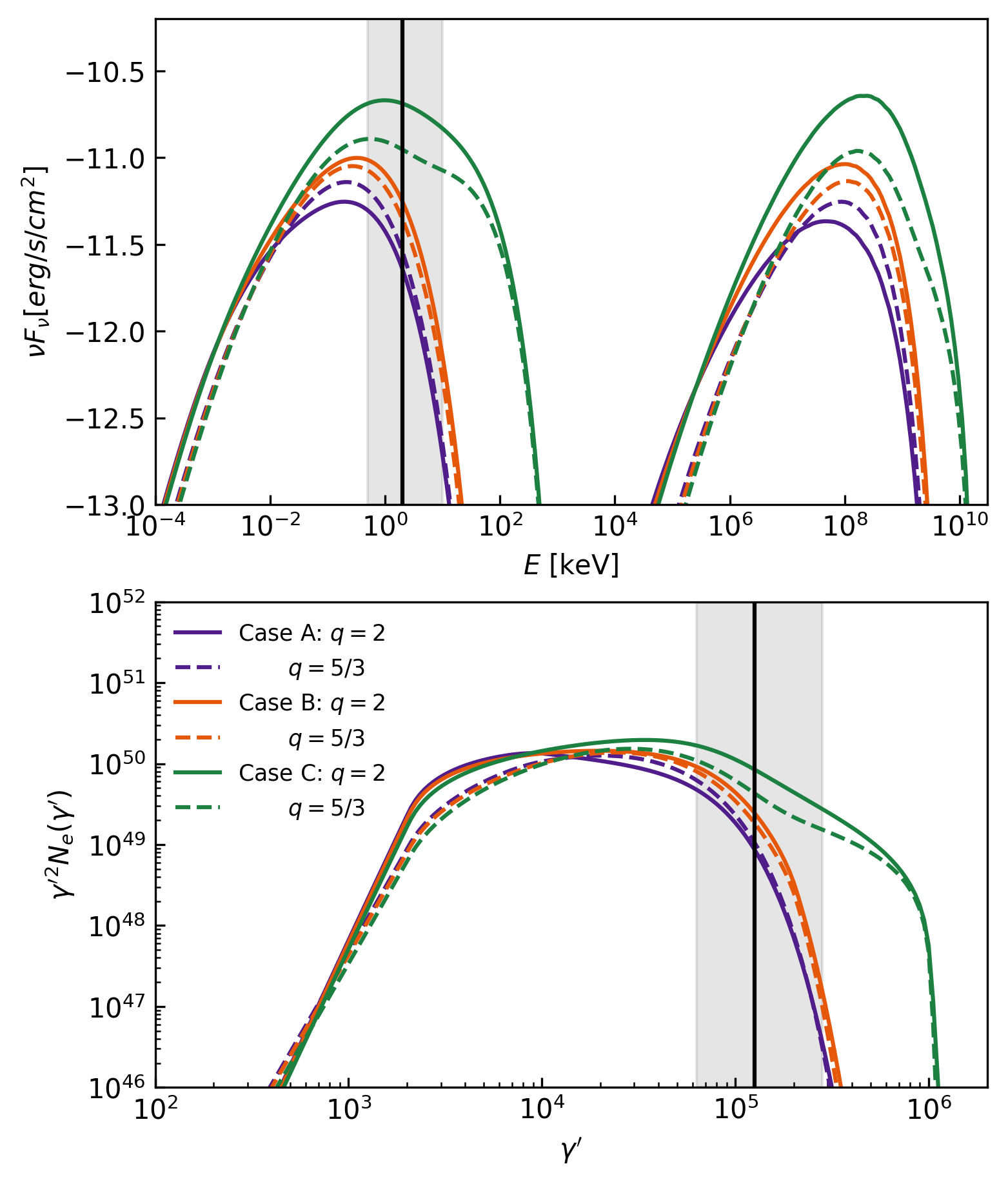}
   \caption{
   Steady-state SEDs and EEDs obtained for the Kolmogorov turbulence index $q = 5/3$.
        }
   \label{figs:sed_q}%
    \end{figure}

Figure \ref{figs:sed_q} displays the steady-state energy spectra of the electrons and photons obtained for the Kolmogorov turbulence index $q=5/3$.
For comparison, the corresponding results for $q=2$ are also presented in the figure.

\section{Influence of the \emph{Fermi}-I acceleration}\label{sec:append3}

\begin{figure*}
   \centering
  \includegraphics[width=0.9\textwidth]{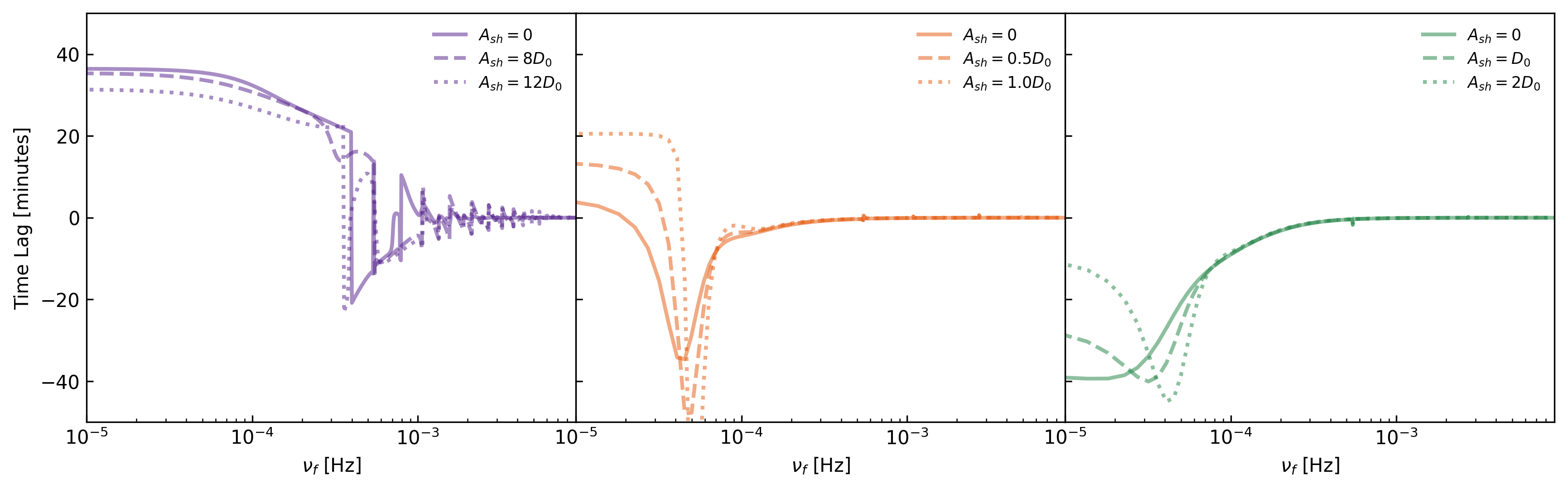}
   \caption{
   Influence of the \emph{Fermi}-I acceleration on time lags for three typical lag regimes.
        }
   \label{figs:lags_q}%
    \end{figure*}

In Fig. \ref{figs:lags_q}, we show the lag–frequency spectra after incorporating the \emph{Fermi}-I acceleration.

\bibliographystyle{apsrev4-2}
\bibliography{references}

\end{document}